\documentclass[preprintnumbers,article,amsmath,amssymb,floatfix,10pt,prd,onecolumn,
superscriptaddress,nofootinbib]{revtex4-2}
\usepackage{bm}
\usepackage{amsfonts}
\usepackage{latexsym}
\usepackage{graphicx}
\usepackage{amsmath}
\usepackage{textcomp}
\usepackage{hyperref}
\hypersetup{colorlinks,citecolor=blue}
\hypersetup{colorlinks=true,linkcolor=red,filecolor=magenta,    urlcolor=blue}
\usepackage{amsmath}
\usepackage{xcolor}\usepackage{cancel, ulem}
\usepackage{epsfig}

\newcommand{\m}{\mathring}
\newcommand{\be}{\begin{equation}}
\newcommand{\ee}{\end{equation}}
\newcommand{\bea}{\begin{eqnarray}}
\newcommand{\eea}{\end{eqnarray}}
\newcommand{\eeas}{\end{eqnarray*}}
\newcommand{\beas}{\begin{eqnarray*}}

\newcommand{\Lim}{\displaystyle\lim}

\def\jnl@style{\it}
\def\aaref@jnl#1{{\jnl@style#1}}

\def\aaref@jnl#1{{\jnl@style#1}}

\def\aj{\aaref@jnl{AJ}}                   
\def\apj{\aaref@jnl{ApJ}}                 
\def\apjl{\aaref@jnl{ApJ}}                
\def\apjs{\aaref@jnl{ApJS}}               
\def\apss{\aaref@jnl{Ap\&SS}}             
\def\aap{\aaref@jnl{A\&A}}                
\def\aapr{\aaref@jnl{A\&A~Rev.}}          
\def\aaps{\aaref@jnl{A\&AS}}              
\def\mnras{\aaref@jnl{Mon.~Not.~Roy.~Astron.~Soc.}}             
\def\prd{\aaref@jnl{Phys.~Rev.~D}}        
\def\prc{\aaref@jnl{Phys.~Rev.~C}}  
\def\prl{\aaref@jnl{Phys.~Rev.~Lett.}}    
\def\qjras{\aaref@jnl{QJRAS}}             
\def\skytel{\aaref@jnl{S\&T}}             
\def\ssr{\aaref@jnl{Space~Sci.~Rev.}}     
\def\zap{\aaref@jnl{ZAp}}                 
\def\nat{\aaref@jnl{Nature}}              
\def\aplett{\aaref@jnl{Astrophys.~Lett.}} 
\def\apspr{\aaref@jnl{Astrophys.~Space~Phys.~Res.}} 
\def\physrep{\aaref@jnl{Phys.~Rep.}}      
\def\physscr{\aaref@jnl{Phys.~Scr}}       
\def\commat{\aaref@jnl{Comm.~Math.~Phys.}}              
\def\science{\aaref@jnl{Science}}               
\def\cqg{\aaref@jnl{Classical Quant.~Grav.}}            
\def\jpcs{\aaref@jnl{JPCS}}                                     
\def\ijmpd{\aaref@jnl{Int.~J.~Mod.~Phys.~D}}                    
\def\grg{\aaref@jnl{Gen.~Relat.~Gravit.}}               
\def\rpp{\aaref@jnl{Rep.~Prog.~Phys.}}          
\def\npa{\aaref@jnl{Nucl.~Phys.~A}}        
\def\lrr{\aaref@jnl{Living Rev.~Rel.}}                   
\def\jcap{\aaref@jnl{J.~Cosmology Astropart.~Phys.}}    
\def\rmp{\aaref@jnl{Rev.~Mod.~Phys.}}   
\def\epjc{\aaref@jnl{Eur.~Phys.~J.~C}} 
\def\plb{\aaref@jnl{~Phy.~Lett.~B}} 
\def\mpla{\aaref@jnl{Mod.~Phy.~Lett.~A}} 
\def\arxiv{\aaref@jnl{arxiv.org}}

\allowdisplaybreaks[1]

\begin{document}

\title{Phase-space analysis of a novel cosmological model in $f(Q)$ theory}

\author{Hamid Shabani}
\email{h.shabani@phys.usb.ac.ir}
\affiliation{Physics Department, Faculty of Sciences, University of Sistan and Baluchestan, Zahedan, Iran}
\author{Avik De}
\email{avikde@utar.edu.my}
\address{Department of Mathematical and Actuarial Sciences\\
Universiti Tunku Abdul Rahman, Jalan Sungai Long, 43000 Cheras, Malaysia}
\author{Tee-How Loo}
\email{looth@um.edu.my}
\affiliation{Institute of Mathematical Sciences, Faculty of Science, Universiti Malaya, 50603 Kuala Lumpur, Malaysia}

\footnotetext{The research was supported by the Ministry of Higher Education (MoHE), through the Fundamental Research Grant Scheme (FRGS/1/2021/STG06/UTAR/02/1). }
\begin{abstract}
The vanishing affine connections have been used solely while adopting the modified $f(Q)$ gravity theory to the cosmology. Consequently, researchers could not get beyond what is already known in $f(T)$ theory earlier. To alleviate this problem, in the present manuscript we investigate a recently proposed construction of $f(Q)$ theory using non-vanishing affine connection in the spatially flat FLRW spacetime. We then investigate the cosmological solutions of $f(Q)$ theory for a perfect fluid through the phase space analysis. We introduce few variables and dimensionless parameters to construct the corresponding equations suitable for the dynamical system approach. The conservation of the energy–momentum tensor leads to a constraint equation that relates the dynamical variables. Briefly, both unstable and stable de Sitter solutions appear which correspond to early and late times accelerated expansions. Also, unstable points corresponding to the matter dominated and radiation dominated eras have been found which do exist for every $f(Q)$ function. As a result, the present discussion shows that $f(Q)$ gravity endowed by non-vanishing affine connections is capable of explaining a true sequence of cosmic eras.
\end{abstract}
\maketitle

\section{Introduction}\label{sec1}
All this while, the underlying spacetime has three geometric entities: the curvature, the torsion and the non-metricity, which either individually or jointly can attribute the gravity. However, only the curvature of the spacetime was used to formulate Einstein's general relativity (GR) in a torsion-free and metric-compatible (or, vanishing non-metricity) environment. The covariant differentiation of tensors were based on the very special and unique affine connection, 
the Levi-Civita connection in this standard theory of gravity. 
In ``Symmetric teleparallel gravity", this special affine connection is replaced by a general affine connection on a flat spacetime with vanishing torsion, offering its non-metricity alone to drive gravity. 
In an earlier ``metric teleparallel theory", which is based on an affine connection with vanishing curvature and non-metricity, Einstein himself \cite{1} ascribed gravity to the torsion scalar $\mathbb{T}$  of spacetime in an attempt to unite gravity with electromagnetism. 
In the currently discussed symmetric teleparallel theory, one considers the Lagrangian density $\mathcal{L}=\sqrt{-g}Q$ using the non-metricity scalar $Q$, to obtain the respective field equations.
A similar treatment developed its metric counterpart. 
However, the latter two theories are equivalent to GR up to a boundary term and naturally also suffer from the dark sector issues. 
As a result, extensions in terms of $f(\mathbb{T})$ and $f(Q)$-theories were formulated in the respective categories, which dynamically diverge from GR.
Of course, the $f(\mathbb{T})$ theory was proposed \cite{fT1st} much earlier and analyzed for a long time theoretically and also in cosmological side \cite{teleparallel} and it is almost at its maturity, whereas the newly proposed \cite{coincident} $f(Q)$ theory is merely at its infancy and a lot of theoretical investigations, viability of functional forms, and contact with observational dataset is still due. 
The interesting comparison of these three modes of gravity theories are well-studied \cite{fQfT,fQfT1,fQfT3}. 
Recently, on a serious inspection on the literature about cosmological applications of $f(Q)$ theories, we notice the tendency of using only the vanishing affine connection in the spatially flat Friedmann-Lema\^itre-Robertson-Walker (FLRW) spacetime, however, the Friedmann equations in this case match exactly with those of the $f(\mathbb{T})$ theory \cite{cosmology_Q}. 
Therefore, this particular gauge bounds the researchers within the results we already obtained in $f(\mathbb{T})$ theory and consequently the importance of $f(Q)$ theory as an emerging novel modified gravity theory is defeated. 
This motivated us to look for a new construction \cite{FLRW/connection} of $f(Q)$ theory based on another gauge equivalence class of (non-vanishing) affine connections involving a parameter $\gamma(t)$ in the spatially flat FLRW background. 

The present article is organized as follows:\\
After this introductory part, in Section \ref{sec2} we set the stage of our present discussion by providing the basic mathematical background of $f(Q)$ theory. The equations of motion of $f(Q)$ gravity in a spatially flat FLRW spacetime whose line element is given in spherical polar coordinates and formulated based on a non-vanishing yet torsion-free and flat affine connection is introduced in Section \ref{sec3}. In the next Section \ref{seccom}, an additional constraint in the form of energy conservation is discussed to construct a closed dynamical system\footnote{To consider a comprehensive study of the application of the autonomous dynamical systems in various modified theories of gravity see Ref.~\cite{bahamonde2018} and also Refs.\cite{odintsov2017,oikonomou2018,odintsov2018,chatzarakis2020}.}, followed by the formation of the actual system of dimensionless variables in Section \ref{sec5}. To elaborate the analysis a pure power-law model is considered in Subsection \ref{var-a}. Next, a constant $\gamma$ is assumed in Section \ref{sec-con} and the corresponding field equation and the dynamical system is analyzed. Three different models are discussed in the subsections~\ref{cons-a},~\ref{cons-b} and~\ref{cons-c}. Finally, we conclude all our findings in Section \ref{sec8}.


\section{Symmetric teleparallel formulation}\label{sec2}
In the symmetric teleparallel theory of gravity, a general affine connection $\Gamma^\lambda{}_{\mu\nu}$ 
with vanishing curvature and null torsion is considered and let its non-metricity property alone control the gravity. We define the non-metricity tensor 
\begin{equation} \label{Q tensor}
Q_{\lambda\mu\nu} = \nabla_\lambda g_{\mu\nu} \,.
\end{equation}
The two types of traces of the non-metricity tensor are 
\[
Q_{\lambda}=Q_{\lambda\mu\nu}g^{\mu\nu}; \quad \tilde Q_{\nu}=Q_{\lambda\mu\nu}g^{\lambda\mu}.
\]
The disformation tensor $L^\lambda{}_{\mu\nu}$ and the superpotential tensor $P^\lambda{}_{\mu\nu}$ are respectively given by
\begin{equation} \label{L}
L^\lambda{}_{\mu\nu} = \frac{1}{2} (Q^\lambda{}_{\mu\nu} - Q_\mu{}^\lambda{}_\nu - Q_\nu{}^\lambda{}_\mu) \,.
\end{equation}
\begin{equation} \label{P}
P^\lambda{}_{\mu\nu} = \frac{1}{4} \left( -2 L^\lambda{}_{\mu\nu} + Q^\lambda g_{\mu\nu} - \tilde{Q}^\lambda g_{\mu\nu} -\frac{1}{2} \delta^\lambda_\mu Q_{\nu} - \frac{1}{2} \delta^\lambda_\nu Q_{\mu} \right) \,.
\end{equation} 
It is well-known that the disformation tensor serves as the linkage between the affine connection and the Levi-Civita one $ \m{\Gamma}^\lambda{}_{\mu\nu}$:
\begin{equation} \label{connc}
\Gamma^\lambda{}_{\mu\nu} = \mathring{\Gamma}^\lambda{}_{\mu\nu}+L^\lambda{}_{\mu\nu}\,.
\end{equation}
We consider non-metricity scalar  
\begin{equation} \label{Q}
Q=Q_{\lambda\mu\nu}P^{\lambda\mu\nu}= \frac{1}{4}(-Q_{\lambda\mu\nu}Q^{\lambda\mu\nu} + 2Q_{\lambda\mu\nu}Q^{\mu\lambda\nu} +Q_\lambda Q^\lambda -2Q_\lambda \tilde{Q}^\lambda).
\end{equation}

However, being equivalent to GR, the symmetric teleparallelism inherit the same `dark' problem as in GR, and so a modified $f(Q)$ gravity has been introduced \cite{coincident} in the same way as a modified $f(R)$ theory was introduced to extend GR. By varying the action term of $f(Q)$ theory  
\begin{equation*}
S = \frac1{2\kappa}\int f(Q) \sqrt{-g}\,d^4 x
+\int \mathcal{L}_M \sqrt{-g}\,d^4 x
\end{equation*}
with respect to the metric we obtain the field equation
\begin{equation} \label{FE1}
\frac{2}{\sqrt{-g}} \nabla_\lambda (\sqrt{-g}FP^\lambda{}_{\mu\nu}) -\frac{1}{2}f g_{\mu\nu} + F(P_{\nu\rho\sigma} Q_\mu{}^{\rho\sigma} -2P_{\rho\sigma\mu}Q^{\rho\sigma}{}_\nu) = \kappa T^{m}_{\mu\nu}.
\end{equation}
where we have defined $F=df/dQ$  (differentiations with respect to the argument will later be shown by primes). We consider the stress-energy tensor $T^{m}_{\mu\nu}$ to be a perfect fluid given by
\begin{align}
T^{m}_{\mu\nu}=(p+\rho)u_\mu u_\nu+pg_{\mu\nu}
\end{align} 
where $p$ and $\rho$ denote the pressure and energy density of the ordinary matter.

Very recently, the covariant formulation of this field equation was obtained and used effectively in cosmological sector \cite{zhao}
\begin{equation} \label{FE}
F \m{G}_{\mu\nu}+\frac{1}{2} g_{\mu\nu} (FQ-f) + 2F' P^\lambda{}_{\mu\nu} \m{\nabla}_\lambda Q = \kappa T^{m}_{\mu\nu}
\end{equation}
where $$\mathring{G}_{\mu\nu} = \mathring{R}_{\mu\nu} - \frac{1}{2} g_{\mu\nu} \mathring{R}.$$ All the expressions with a $\mathring{()}$ is calculated with respect to the Levi-Civita Connection. Hence in its GR equivalent form we can express it as \cite{de/comment}
\begin{align}\label{equiv}
    \mathring{G}_{\mu\nu}=\frac{\kappa}{F}\mathcal{T}_{\mu\nu}=\frac{\kappa}{F}T^{m}_{\mu\nu}+\kappa T^{\text{DE}}_{\mu\nu},
\end{align}
where
\begin{align*}
\kappa T^{\text{DE}}_{\mu\nu}=\frac{1}{F}\left[\frac{1}{2}g_{\mu\nu}(f-QF)-2F'\mathring{\nabla}_\lambda QP^\lambda_{\mu\nu}\right].
\end{align*}
Several important publications came up very recently on this modified $f(Q)$-gravity theory and its cosmological implications, see \cite{de/comment, cosmology, cosmography, lin, de/isotropization, de/accelerating, de/complete, redshift, signature, lcdm1, siren, anagnostopoulos2021, lazkos2019, arora2022, gadbail2022, harko2018, bajardi2020, lymperis2022, sahoo2022, aziza2021, de2022} and the references therein. The dynamical system analysis of the $f(Q)$ theory in both background and perturbative level of a spatially flat FLRW spacetime were also conducted \cite{dyn1,dyn2,dyn3,dyn4}. However, all these studies were conducted in the coincident gauge choice, line element in Cartesian coordinates together with vanishing $\Gamma^\lambda{}_{\mu\nu}$. This specific choice reduced the covariant derivative into partial derivative, making the calculations simpler. 
But at the adverse side, the Friedmann-like energy and pressure equations are identical with the $f(T)$ theory.

Varying the action term with respect to the affine connection, we obtain the other field equation of $f(Q)$ theory, on the basis of the assumption that the matter Lagrangian $\mathcal{L}_{M}$ is not a function of the affine connection, 
\begin{align}\label{FE2}
\nabla_\mu\nabla_\nu(\sqrt{-g}F P^{\nu\mu}{}_\lambda)=0\,.
\end{align}


\section{The homogeneous and isotropic model of the universe}\label{sec3}

The spatially flat homogeneous and isotropic FLRW spacetime metric is given by
\begin{align}\label{metric}
g = -\mathrm{d} t\otimes \mathrm{d} t + a\left(t\right)^{2}\left(   \mathrm{d} r\otimes \mathrm{d} r + r^{2}  \mathrm{d} {\theta}\otimes \mathrm{d} {\theta} + r^{2} \sin\left({\theta}\right)^{2} \mathrm{d} {\phi}\otimes \mathrm{d} {\phi}\right)\,.
\end{align}
In the current discussion we consider a non-vanishing class of affine connections $\Gamma$, first introduced and studied from the cosmological perspective in \cite{FLRW/connection}, which are torsion free with zero curvature, yet non-compatible with the metric (\ref{metric})
\begin{align} \label{gamma1}
\Gamma^t{}_{tt}&=\gamma+\frac{\dot{\gamma}}{\gamma}, 
    \quad       \Gamma^r{}_{tr}=\gamma, 
	\quad		\Gamma^r{}_{\theta\theta}=-r, 
	\quad		\Gamma^r{}_{\phi\phi}=-r\sin^2\theta,												
    \quad       \Gamma^\theta{}_{t\theta}=\gamma, 
	\notag\\
        		\Gamma^\theta{}_{r\theta}&=\frac1r,
	\quad		\Gamma^\theta{}_{\phi\phi}=-\cos\theta\sin\theta,			\quad   \Gamma^\phi{}_{t\phi}=\gamma, 
	\quad 	\Gamma^\phi{}_{r\phi}=\frac1r, 
	\quad 	\Gamma^\phi{}_{\theta\phi}=\cot\theta,
\end{align}
where $\gamma(t)$ is a nonvanishing function of $t$, so far unconstrained by theoretical results or observational data. Also, an overdot indicates a time derivative. It produce nonzero non-metricity tensor components which provides us novel insight of the $f(Q)$ dynamics in the spatially flat FLRW background, keeping it completely aloof from the $f(T)$ theory dynamics. However, due to the constraint of energy conservation, or equivalently the connection field equation (\ref{FE2}), we pay the price with a significant restrictions on the viable choices of the functional form $f$ \cite{ad/bianchi}.
 The non-metricity scalar $Q$ can be computed as
\begin{align}\label{Q-2}
    Q=&-6H^2+9\gamma H+3\dot\gamma.         
\end{align}
The Friedmann-like equations corresponding to the field equation (\ref{FE}) are given by
\begin{align}
 \kappa \rho=&\frac12f+\left(3H^2-\frac Q2\right)F +\frac32\dot{Q}\gamma F'\,,\label{rho}\\
 \kappa p=&-\frac12f+\left(-2\dot{H}-3H^2+\frac Q2 \right)F +\frac{\dot{Q}}{2}\left(-4H+3\gamma\right)F'\,.
            \label{p}
\end{align}

 The divergence of the energy-momentum tensor $T_{\mu\nu}$ \cite{ad/bianchi} yields the continuity relation 
   \begin{align}\label{cr}
    \kappa \dot{\rho}+3H\kappa(p+\rho)=\frac32\gamma[(\ddot{Q}+3H\dot{Q})F'+\dot{Q}^2F''].
 \end{align}
%
%
\section{Comments on the continuity relation}\label{seccom}

Before analyzing the equations~(\ref{rho}) and~(\ref{p}) via the dynamical system approach we briefly discuss the continuity relation~(\ref{cr}). In the next sections we will define a closed dynamical system corresponding to the equations~(\ref{rho}) and~(\ref{p}) based on the conservation of the ordinary matter stress-energy tensor assumption, i.e., $\mathring{\nabla}^{\mu} T^{m}_{\mu \nu}=0$. From the equation~(\ref{cr}) this leads to 
\begin{align}\label{cr1}
\ddot{F}+3H\dot{F}=0,
\end{align}
with the following solution
\begin{align}\label{cr2}
F=C\int a^{-3} dt+D,
\end{align}
where $C$ and $D$ being some constants and $a$ is the scale factor. One observes that the equations~(\ref{Q-2}) and~(\ref{cr2}) for a known function $f(Q)$ can lead to a first order differential equation for $\gamma$ function. For example for $f(Q)=Q^{n}$ one obtains
\begin{align}\label{cr3}
3\dot{\gamma} + 9\gamma H-6H^2=\left[\frac Cn\int a^{-3}dt+\frac Dn\right]^{\frac1{n-1}}.
\end{align}
Note that only two of the equations (\ref{rho})-(\ref{cr}) are independent. This means that without assuming the conservation of the matter stress-energy tensor we have only two independent equations for three unknowns $\rho$, $a$ and $\gamma$ (considering $p=w\rho$ for the ordinary matter). Therefore, at least from the mathematical point of view by assuming the conservation of matter stress-energy tensor we already get a solution for one of the variables, i.e., $\rho\propto a^{-3(1+w)}$. In this case equations~(\ref{cr2}) and~either of~(\ref{rho}) or~(\ref{p}) can be used to determine the remaining variables.


\section{Field equations in terms of dimensionless variables}\label{sec5}

In the present section, we rewrite the field equations~(\ref{rho})-(\ref{cr}) in terms of some dimensionless variables. To this end, we suppose that the stress-energy tensor of the ordinary matter is conserved. This gives the equation~(\ref{cr1}). Thus, we begin with the following definitions  

\begin{align}\label{ds1}
&x_1=-\frac{f}{6H^{2}F},~~~~~x_2=\frac{Q}{6H^{2}},~~~~~x_3=-\frac{\dot{Q}F'}{HF}=-\frac{\dot{F}}{HF},~~~~~x_4=\frac{\gamma}{2H},~~~~~
    \Omega^{m} =\frac{\kappa\rho}{3H^{2}F},\nonumber\\
&r=-\frac{QF}{f}=\frac{x_2}{x_1},~~~~~~m=\frac{QF'}{F}.
\end{align}

As we will see, the parameters $r$ and $m$ can parametrize the function $f(Q)$. In fact, by eliminating $Q$ from the definitions of $m$ and $r$ one gets a function $m(r)$, at least for well-behaved $f(Q)$ functions. Rewriting the equations~(\ref{rho}), (\ref{p}) and (\ref{cr1}) in terms of the above variables leads to

\begin{align}
&\Omega^{m} =1-x_1-x_2-x_3x_4,\label{ds2}\\
&\frac{2\dot{H}}{3H^{2}}=-1-w\Omega^m+x_1+x_2+x_3(\frac23-x_4),\label{ds3}\\
&\frac{\ddot{F}}{H^{2}F}=3x_3\label{ds4},
\end{align}
respectively. Here, one can define an ``effective equation of state parameter" as it is usual in the modified theories of gravity. In this regard, we obtain

\begin{align}\label{ds5}
w^{eff}=-1-\frac{2\dot{H}}{3H^{2}}=w\Omega^m-x_1-x_2-x_3(\frac23-x_4).
\end{align}

Comparing to the GR field equations, by defining an effective equation of state, we consider an effective stress-energy tensor. In fact, this effective tensor is defined as $T_{\mu\nu}^{eff}=\mathcal{T}_{\mu\nu}/F$ (see the equation (\ref{equiv})) which represents the effective equivalent form of the GR equations. The first order dynamical equations are obtained as

\begin{align}
&\frac{dx_1}{dN}=\frac{x_2 x_3}{m}-3 x_1 \left[w \left(x_1+x_2+x_3 x_4-1\right)+x_1+x_2+x_3\left(\frac{ 1}{3}-x_4\right)-1\right],\label{ds6}\\
&\frac{dx_2}{dN}=-3x_2\left[w \left(x_1+x_2+x_3 x_4-1\right)+x_1+x_2+x_3\left(\frac{1}{3m}+\frac{2}{3}- x_4\right)-1\right],\label{ds7}\\
&\frac{dx_3}{dN}=-\frac{3}{2} x_3 \left[w \left(x_1+x_2+x_3 x_4-1\right)+x_1+x_2-x_3 x_4+1\right],\label{ds8}\\
&\frac{dx_4}{dN}=-\frac{3}{2} x_4 \left[w \left(x_1+x_2+x_3 x_4-1\right)+x_1+x_2+x_3\left(\frac{2}{3}-x_4\right)+1\right]+x_2+1,\label{ds9}
\end{align}
where to obtain the equations~(\ref{ds8}) and~(\ref{ds9}) the expressions~(\ref{ds4}) and~(\ref{Q-2}) have respectively been used. The system~(\ref{ds6})-(\ref{ds9}) is not autonomous since $m=m(r)=m(x_2/x_1)$ has been appeared in Eqs.~(\ref{ds6}) and~(\ref{ds7}). Hence, without specifying the form of $f(Q)$ function one cannot solve the system. In this regard, in the subsections~\ref{var-a} and~\ref{cons-a} we discuss models with $m=constant$ and two cases with particular form of $m(r)$ parameter will be investigated in the subsections~\ref{cons-b} and~\ref{cons-c} when $\gamma=constant$. In the case of a varying $\gamma$ parameter there are four independent variables and thus we are facing a four dimensional phase space. Different models can be investigated via Eqs.~(\ref{ds6})-(\ref{ds9}). However, in this section we only consider a power-low form for which the phase space reduces to a three dimensional one and thus we can present a better illustration of the phase space. When $\gamma$ is constant the set of equations get three dimensions, since, one is capable of demonstrating the whole phase space. We postpone considering the effects of a constant $\gamma$ to the section ~\ref{sec-con}. Note that, throughout the paper we consider only $w=0$, although for being complete all equations of motion are obtained for general $w$.


%
\subsection{$f(Q)=\alpha Q^{\beta}$ gravity with constant values of $\alpha$ and $\beta$} \label{var-a}

We consider the conditions under which acceptable cosmological solutions do exist. An admissible solution describes a transition from an unstable de Sitter era to the matter dominated era followed by a stable dark energy dominated one. In the case of the power-law model it can be obtained
\begin{align}\label{case1-1}
m=\beta-1,~~~~~~r=-\beta.
\end{align}
The result~(\ref{case1-1}) means that $m$ and $r$ both are constants. The corresponding dynamical system has three dimensions since $x_{2}=rx_{1}$ in this case, for which the critical points have been presented in Table~\ref{tab11}. 
\begin{center}
\begin{table}[h]
\centering
\caption{The fixed points solutions of $f(Q)=\alpha Q^{\beta}$ gravity with variable $\gamma$.}
\begin{tabular}{l @{\hskip 0.1in} l@{\hskip 0.1in} l @{\hskip 0.1in}l @{\hskip 0.1in}l}\hline\hline
Fixed point     &Coordinates $(x_1,x_3,x_4)$           &Eigenvalues  &$\Omega^m$      &$w^{eff}$\\[0.5 ex]
\hline
$P^a$&$\left(0,\frac{1}{x_4},x_4\right)$&$\left[0,3(1-w)-\frac{1}{x_4},6-\frac{\frac{1}{m}+2}{x_4}\right]$&$0$&$1-\frac{2}{3x_4}$\\[0.75 ex]
$P^{b}$&$\left(0,\frac{6 m}{2 m+1},\frac{1}{6} \left(\frac{1}{m}+2\right)\right)$&$\left[0,0,\frac{3}{2 m+1}-3 w\right]$&$0$&$\frac{2}{2 m+1}-1$\\[0.75 ex]
$P^{m}$&$\left(0,0,\frac{2}{3(1-w)}\right)$&$\left[\frac{3 (w-1)}{2},\frac{3 (w-1)}{2},3 (w+1)\right]$&$1$&$w$\\[0.75 ex]
$P^{ds}$&$\left(-\frac{1}{m},0,\frac{1}{3} \left(\frac{1}{m}+2\right)\right)$&$\left[-3,-3,-3 (w+1)\right]$&$0$&$-1$\\[0.75 ex]

\hline\hline
\end{tabular}
\label{tab11}
\end{table}
\end{center}

It is worth mentioning that Eqs.~(\ref{Q-2})-(\ref{cr}) imply that there are four unknown variables, $\gamma$, $\rho$ (assuming $p=w\rho$), $H$ (or $a$) and $f$, from which only three of them can be independent. However, when one rewrites the equivalent dynamical system those terms which include different types of derivatives define new variables, for instance, consider $x_2$ which include $\dot{\gamma}$ via $Q$ and $x_{3}$ which contains $\dot{F}/F$. Hence, the number of independent variables may generally exceed those of the original equations. This fact, sources the appearance of $x_4$ in Table~\ref{tab11}. Here, by fixing the forms of $\gamma$ and $f$ the number of independent variables match the number of equations and thus the coordinates do not appear in the solutions, see for example Table~\ref{tab3}. Similar problem is observed in~\cite{paliathanasis2023}. Now, we proceed to mention the properties of the solutions of Table~\ref{tab11} which are as follows.

\begin{itemize}
\item \textbf{ The Point  $P^{m}$:}  as in Table~\ref{tab11} is indicated, it is an unstable point for $w>-1$.

\item \textbf{ The Point  $P^{ds}$:} we have a stable de Sitter solution for $w>-1$, whose location in the phase space depends on $m$.

\item \textbf{ The Point  $P^{a}$:} it can display the role of a stable point for the either of the following conditions.
\begin{align}\label{case1-2}
\left\{
\begin{array}{l}
m>0,~~~~~~~~~~ 0<x_4<\frac{1}{3},\\\\
m < -\frac{1}{2},~~~~~~~~~~ 0<x_4<\frac{2 m+1}{6 m}.
\end{array}
\right.
\end{align}
Therefore, only for $0<x_4<\frac{1}{3}$ the point $P^{a}$ shows the features of an accelerated expansion cosmic epoch with $w^{eff}<-1/3$\footnote{$\Lim_{m\to\infty} \frac{2 m+1}{6 m}=\frac13$.}. An interesting point is that because there is an unstable area between $P^{m}$ with $x_4=2/3$ (while $w=0$ is set) and the interval $0<x_4<\frac{1}{3}$, there is no way to exist a direct connection from $P^{m}$ to $P^{a}$. In this sense, we call such a fixed point as ``isolated fixed point". Since, for initial values in the vicinity of $P^{a}$ we have a Universe with a single evolutionary step; for appropriate initial values the Universe evolves to an accelerated expansion state without experiencing enough lasting in a matter dominated one. Note that, one of the eigenvalues of $P^{a}$ is zero, which implies that $P^{a}$ denotes a line of equilibrium for the intervals~(\ref{case1-2}) instead of a single one. The eigenvector corresponding to the zero eigenvalue is obtained as $(0,1)$ in the $(x_1,x_4)$ plane which means that the $x_4$ axis is the line of equilibrium for the values of $x_4$ which are mentioned in~(\ref{case1-2}). In Fig.~\ref{fig11} the behavior of phase space trajectories in the vicinity of $P^{m}$ are drawn for the model with $\beta=-2$\footnote{$X_{i}$'s are related to $x_{i}$'s when a two dimensions part of the phase space is mapped to a unit radius circle.}. To illustrate the interesting feature of $P^{a}$, we have chosen $x_3=1/x_4$. In this case, since one has $x_3=0$ for both $P^{m}$ and $P^{ds}$, none of them can be pictured, suitably. One observes that only near points to the line of equilibrium can be attracted to it, otherwise, they recede.

\begin{figure}[h!]
\begin{center}
\epsfig{figure=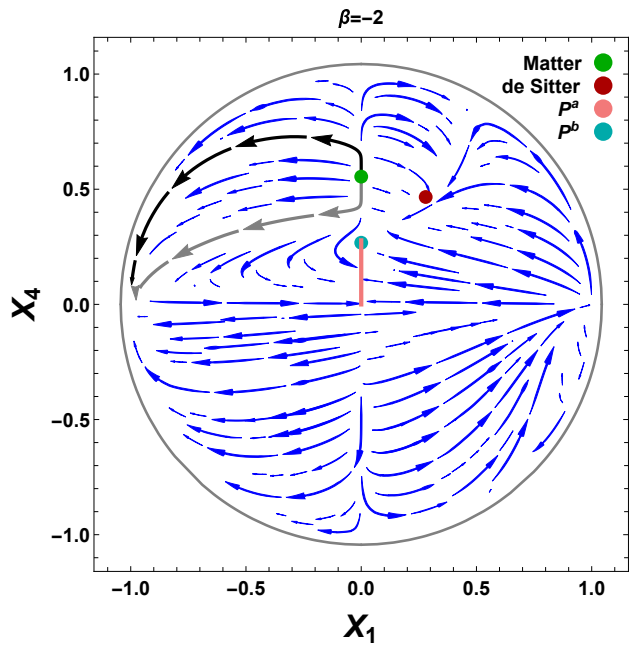,width=8.7cm}
\caption{Illustration of the effect of $P^{a}$ in the plane $(X_1,X_4)$ for a power-law model with $\beta=-2$ when $\gamma$ is a variable. It can be seen that the corresponding line of $P^{a}$ is an attractor only for nearby points. This makes $P^{m}$ a nonconnectable point to $P^{a}$. Note that, because for the best demonstration of the equilibrium line of $P^{a}$ we have chosen $x_3=1/x_4$, the attractive feature of $P^{ds}$ cannot be depicted, truly.}\label{fig11}
\end{center}
\end{figure}

\item \textbf{ The Point  $P^{b}$:} this point is an attractive fixed point which lies within the lower intervals of the constraints~(\ref{case1-2}). It is indicated by a cyan point in Fig.~\ref{fig11}. 

\end{itemize}

\begin{figure}[h!]
\begin{center}
\epsfig{figure=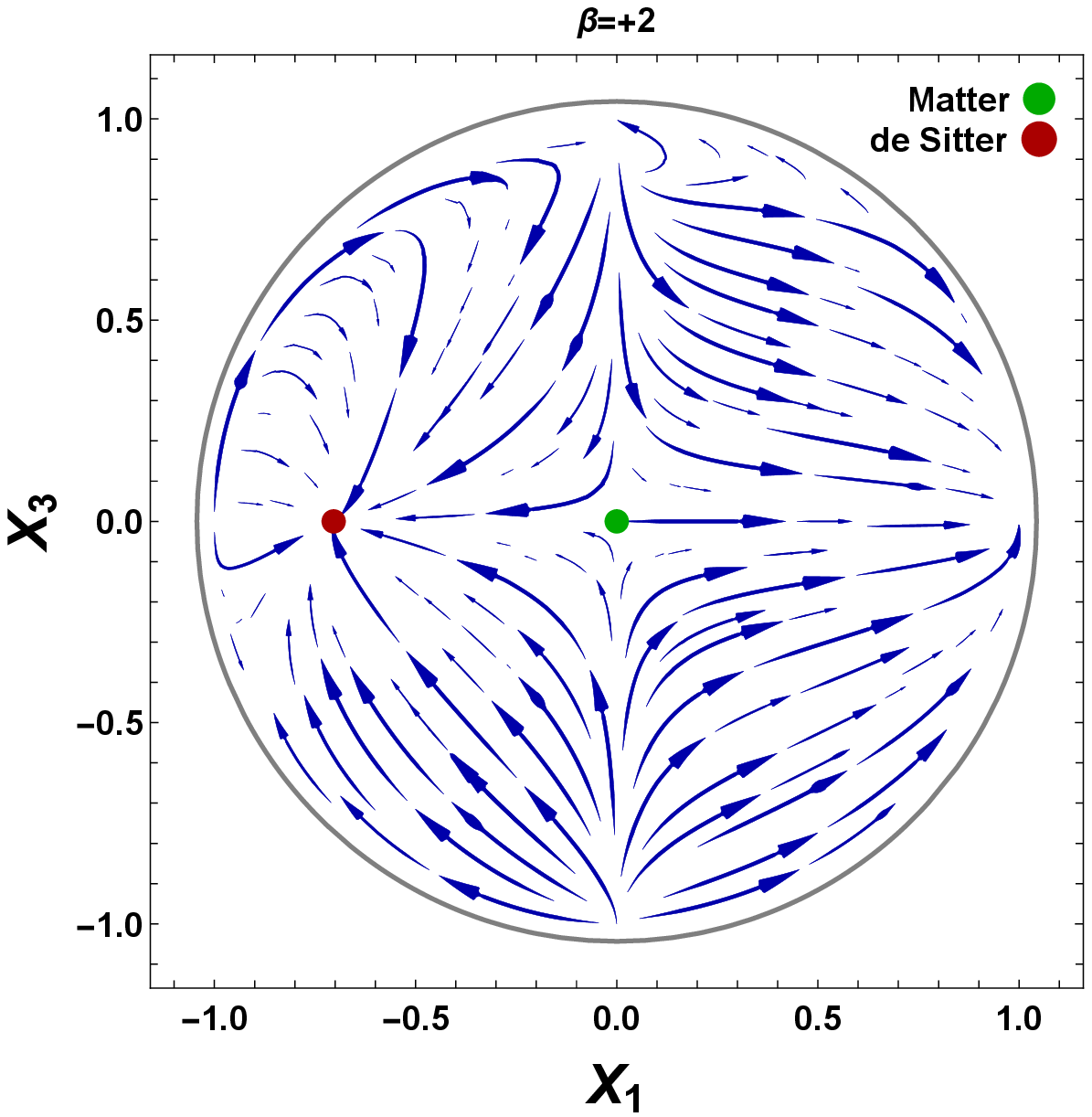,width=8.7cm}\hspace{2mm}
\epsfig{figure=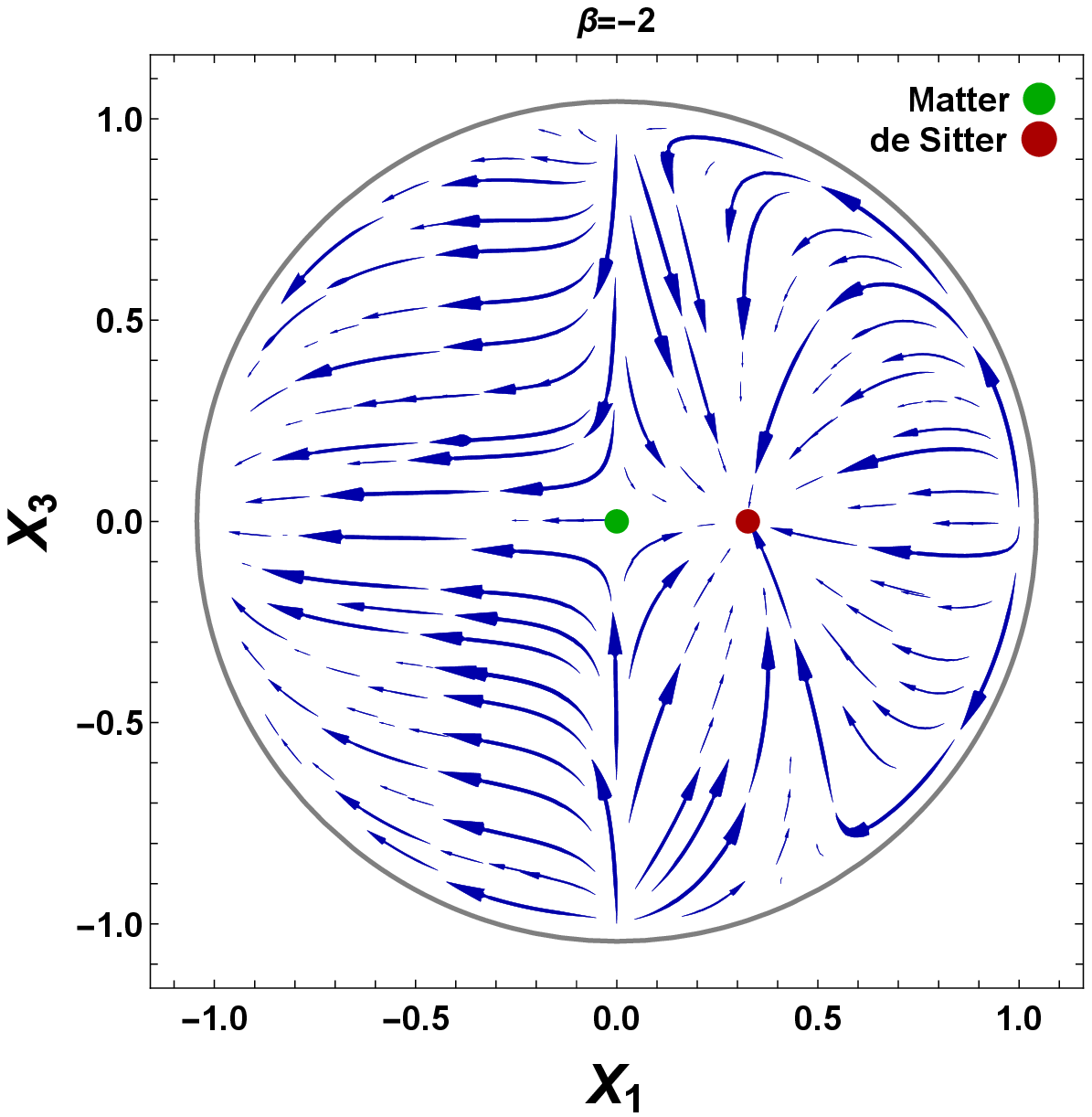,width=8.7cm}
\caption{Phase space portraits of $f(Q)=\alpha Q^{\beta}$ gravity for two values $\beta=\pm2$ when $\gamma$ is a variable. The plane $(x_1,x_3)$ has been selected, because, the two fixed points include $x_3=0$ in their coordinates. Also, we used $x_4=(2m+1)/(3m)$ which is the fourth coordinate of $P^{ds}$. }\label{fig1}
\end{center}
\end{figure}

In Fig.~\ref{fig1} the phase space diagrams of the present model are drawn in the $(X_1,X_3)$ plane ($x_2$ is absent in this model). We have only specified the important points $P^{m}$ and $P^{ds}$ in Fig.~\ref{fig1}. Two models with $\beta=\pm2$ have been selected. As can be seen, the mentioned features are clearly appeared.

\begin{figure}[h!]
\begin{center}
\epsfig{figure=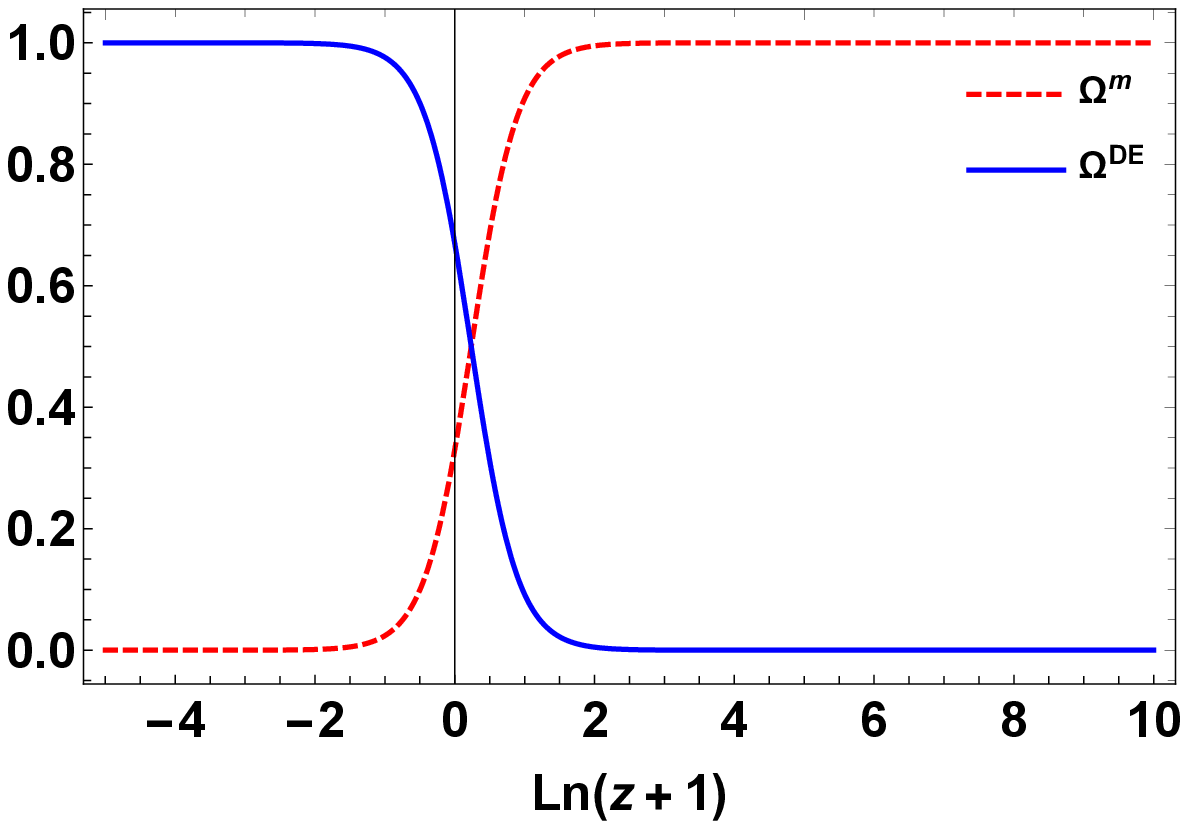,width=8.7cm}\hspace{2mm}
\epsfig{figure=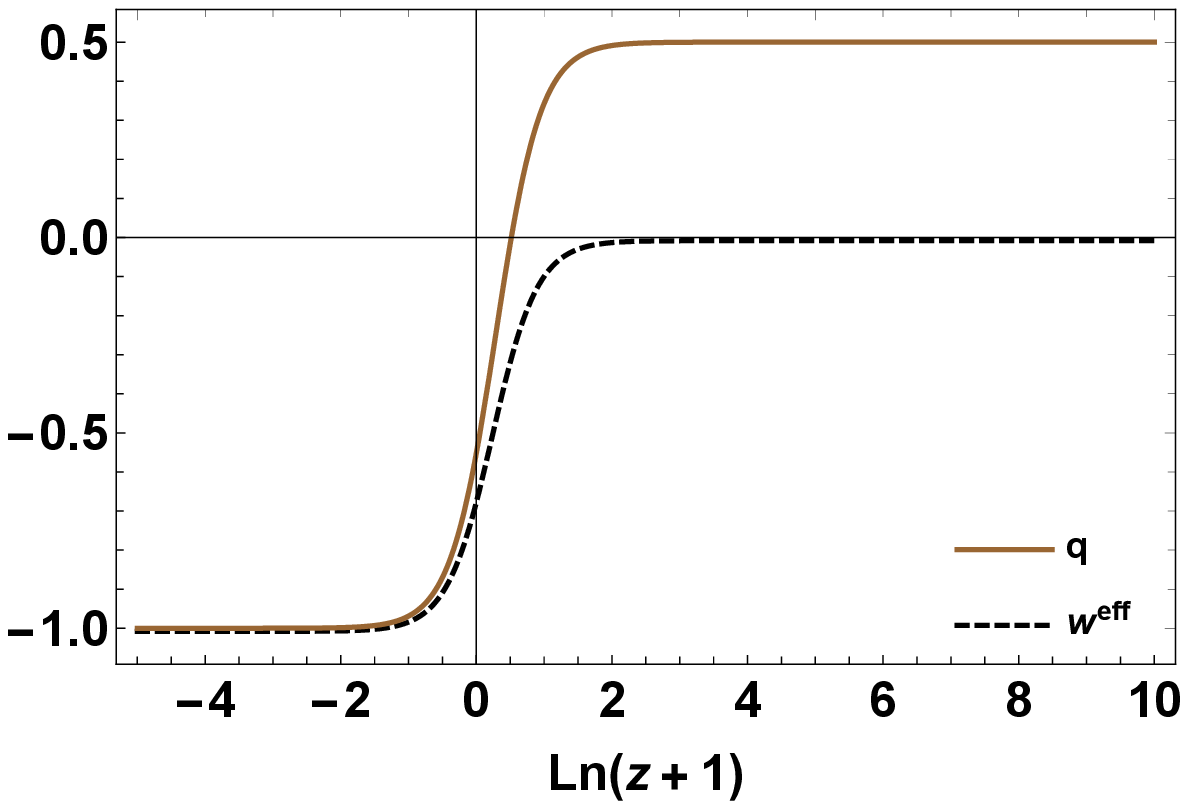,width=8.7cm}\vspace{2mm}
\epsfig{figure=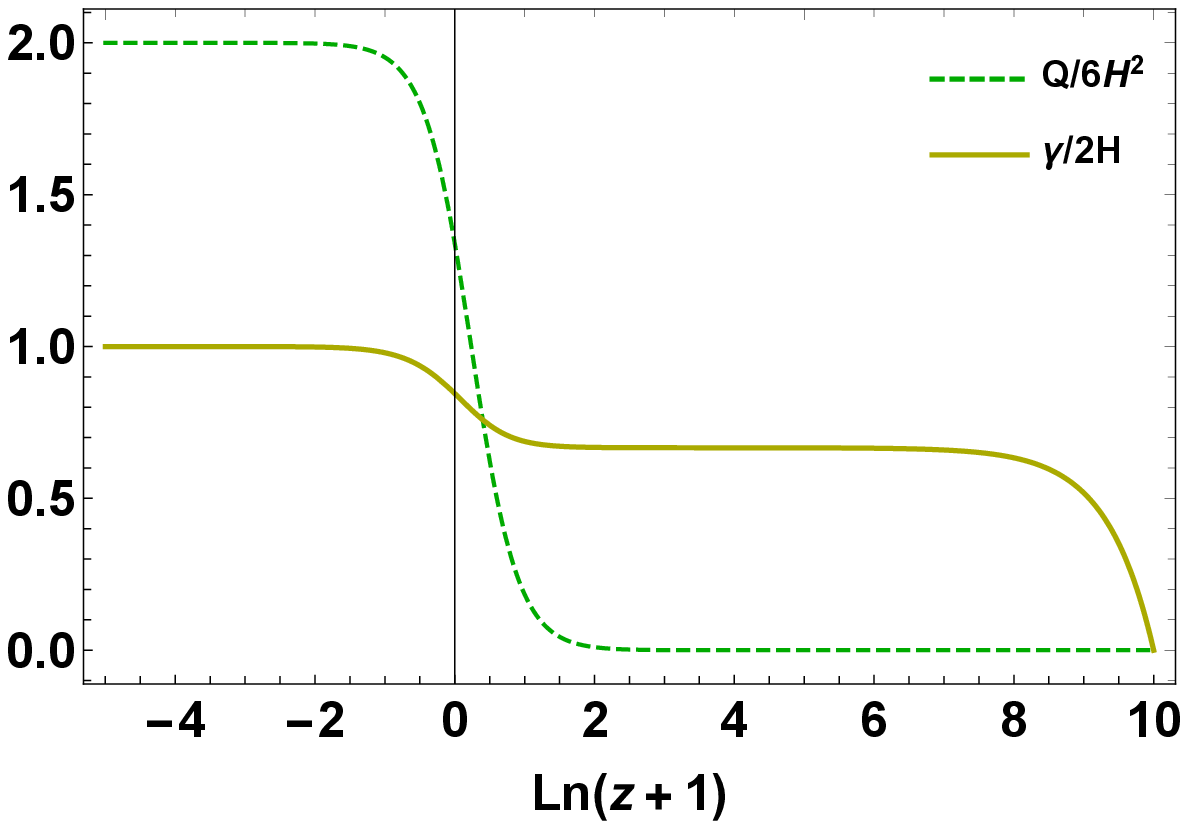,width=8.7cm}
\caption{Cosmological quantities of $f(Q)=\alpha Q^{\beta}$ gravity for $\beta=2$ when $\gamma$ is not constant. We have used initial values $x_{1i}=-2.1\times10^{-13}$, $x_{3i}=1.5\times10^{-15}$ and $x_{4i}=3\times10^{-10}$.}\label{fig2}
\end{center}
\end{figure}

Also, the plots of some important cosmological quantities for $\beta=2$ are depicted in Fig.~\ref{fig2}. They have provided in such a way that the current values of the matter density and the deceleration parameter predicted by this model, match the observations, i.e., to have $\Omega^m_0\approx0.31$ and $q_0\approx-0.55$~\cite{planck2018}. One observes that the deceleration parameter provide an observationally consistent current value. Besides, the lower panel of Fig.~\ref{fig2} indicates that both $\gamma$ parameter and $Q$ are positive valued functions. 

%
%
\section{Field equations in terms of dimensionless variables: The constant $\gamma$ case}\label{sec-con}

Using the same definition given in~(\ref{ds1}) the following changes are made in the case of a constant $\gamma$
\begin{align}
&Q=-6H^{2}+9\gamma H,\label{con-g-1}\\
&x_{4}=\frac13\left(1+x_{2}\right),\label{con-g-2}\\
&\Omega^m=1-x_{1}-x_{2}-\frac13x_{3}\left(1+x_{2}\right),\label{con-g-3}\\
&w^{eff}=-1-\frac{2\dot{H}}{3H^{2}}=w\Omega^m-x_1-x_2-\frac13 x_{3}\left(1-x_{2}\right).\label{con-g-4}
\end{align}

In this case, we reach at a new set of dynamical equations which read

\begin{align}
&\frac{dx_1}{dN}=-3 (w+1) x_1 \Big(x_1+x_2-1\Big)+x_3\left[\frac{x_2}{m}-x_1 \Big((w+1) x_2+w-2\Big)\right],\label{con-g-41}\\
&\frac{dx_2}{dN}=x_2 \bigg[-\frac{x_3}{m}-3 (w+1) \left(x_1+x_2-1\right)-x_3 \Big((w+1) x_2+w-1\Big)\bigg],\label{con-g-42}\\
&\frac{dx_3}{dN}=\frac{1}{2} x_3 \bigg[-3 \Big(w \left(x_1+x_2-1\right)+x_1+x_2+1\Big)-x_3 \Big((w+1) x_2+w-3\Big)\bigg],\label{con-g-43}
\end{align}
where, due to relation~(\ref{con-g-2}), the corresponding equation for $x_4$ is absent. Hence, we have a three dimensional phase space for a general $m$ parameter. The above system can be closed by specifying $m$ in terms of $x_1$ and $x_2$ (see definitions for $r$ and $m$ in~(\ref{ds1})). We proceed with three different models including the case $m=constant$ which corresponds to the power-law models.

\subsection{Models with $f(Q)=\alpha Q^{\beta}$}\label{cons-a}
In the particular case of models with constant 
$m$, i.e., the case of $f(Q)=\alpha Q^{\beta}$ models, the number of independent variables does reduce. In fact, the corresponding equation to $x_2$ is absent because of $x_2=-rx_1$. Also, we add the contribution to the ultra-relativistic matter to the field equations by defining the related energy contribution as

\begin{align}\label{rad1}
\Omega^{rad}=\frac{\kappa\rho^{rad}}{3H^{2}F},
\end{align}

where the superscript ``rad" stands for the ultra-relativistic fluid. In this case, equations~(\ref{ds2}) and~(\ref{ds3}) are changed to

\begin{align}
&\Omega^m =1+ m x_1+\frac{x_3}{3}\Big((m+1) x_1-1\Big)-\Omega^{rad},\label{rad2}\\
&\frac{2\dot{H}}{3H^{2}}=\frac{1}{3} \Bigg[(w-1) x_3-(w+1) x_1 \Big((m+1) x_3+3 m\Big)+3 w \big(\Omega^{rad}-1\big)-\Omega^{rad}-3\Bigg],\label{rad3}
\end{align}

which give

\begin{align}
&\frac{dx_1}{dN}=x_1 \Big((1-3 w) \Omega ^{rad}+3 (w+1)\Big)+\frac{m (1-w)-1}{m}x_3 x_1+ (w+1) x_1^{2} \Big((m+1) x_3+3 m\Big),\label{rad4}\\
&\frac{dx_3}{dN}=\frac{x_3}{2}  \Bigg[(w+1) x_1 \Big((m+1) x_3+3 m\Big)+(1-3 w)\Omega ^{rad}-(w-3) x_3+3( w-1)\Bigg],\label{rad5}\\
&\frac{d\Omega^{rad}}{dN}=\Omega ^{rad} \Bigg[(w+1) x_1 \Big((m+1) x_3+3 m\Big)-(3 w-1) \left(\Omega ^{\text{rad}}-1\right)-(w-2) x_3\Bigg].\label{rad6}
\end{align}

The critical points of the system~(\ref{rad4})-(\ref{rad5}) are shown in Table~\ref{tab3}. Regarding the necessary criteria only solutions $Q^{m}$, $Q^{ds}$ and $Q^{rad}$ imply physical meanings with right stability conditions; $Q^{rad}$ which is the corresponding point of the ultra-relativistic fluid is always unstable. Points $Q^{m}$ and $Q^{ds}$ are unstable and stable for $w>-1$, respectively. Clearly, $Q^a$ is an unstable fixed point by setting $w>-1$. For $Q^b$, the functions $e_1(m,w)$ and $e_2(m,w)$ never simultaneously get negative real values. Therefore, $Q^b$ cannot attract trajectories in the 3D phase space. As can be seen, relation $j(w)>w$ is always valid, and thus, $Q^c$ is also an unstable solution. As a result, the only attractor of the phase space is $Q^{ds}$ which does exist for every value of $m$.
\begin{center}
\begin{table}[h!]
\centering
\caption{The fixed points solutions of $f(Q)=\alpha Q^{\beta}$ gravity with constant $\gamma$.}
\begin{tabular}{l @{\hskip 0.1in} l@{\hskip 0.1in} l @{\hskip 0.1in}l @{\hskip 0.1in}l@{\hskip 0.1in}l}\hline\hline

Fixed point     &Coordinates $(x_1,x_3,\Omega^{rad})$           &Eigenvalues  &$\Omega^m$ &$\Omega^{rad}$     &$w^{eff}$\\[0.5 ex]
\hline
$Q^a$&$\left(0,3+\frac{6}{w-3},0\right)$&$\left[\frac{3}{2} (1-w),\frac{w+3}{3-w},\frac{3 (4 m+w-1)}{m (3-w)}\right]$&$\frac{2}{3-w}$&0&$\frac{w+1}{3-w}$\\[1 ex]
$Q^{b}$&$\left(\frac{1-4 m-w}{m (4 m+3) (w+1)},\frac{6 m}{2 m+1},0\right)$&$\left[\frac{3}{2 m+1}-1, e_1(m,w),e_2(m,w)\footnote{$Q^{b}$ is not implying any physical content, since, $e_{1,2}(m,w)$ functions have not been shown.}\right]$&$\frac{2(1-2 m)}{(w+1)(1+2 m)}$&$0$&$\frac{2}{2 m+1}-1$\\[1 ex]
$Q^c$&$\left(0,2,\frac{w+3}{3 w-1}\right)$&$\left[2-\frac{2}{m},j(w)-w,-(j(w)\footnote{$j(w)=\sqrt{w^2+w+3}.$}+w)\right]$&$\frac{10}{3(1-3 w)}$&$\frac{w+3}{3 w-1}$&$\frac{1}{3}$\\[1 ex]
$Q^{m}$&$\left(0,0,0\right)$&$\left[\frac{3 (w-1)}{2},3 w-1,3 (w+1)\right]$&$1$&0&$w$\\[1 ex]
$Q^{ds}$&$\left(-\frac{1}{m},0,0\right)$&$\left[-4,-3,-3 (w+1)\right]$&$0$&0&$-1$\\[0.75 ex]
$Q^{rad}$&$\left(0,0,1\right)$&$\left[4,-1,(1-3 w)\right]$&$0$&1&$\frac13$\\[0.75 ex]
\hline\hline
\end{tabular}
\label{tab3}
\end{table}
\end{center}

\begin{figure}[ht!]
\begin{center}
\epsfig{figure=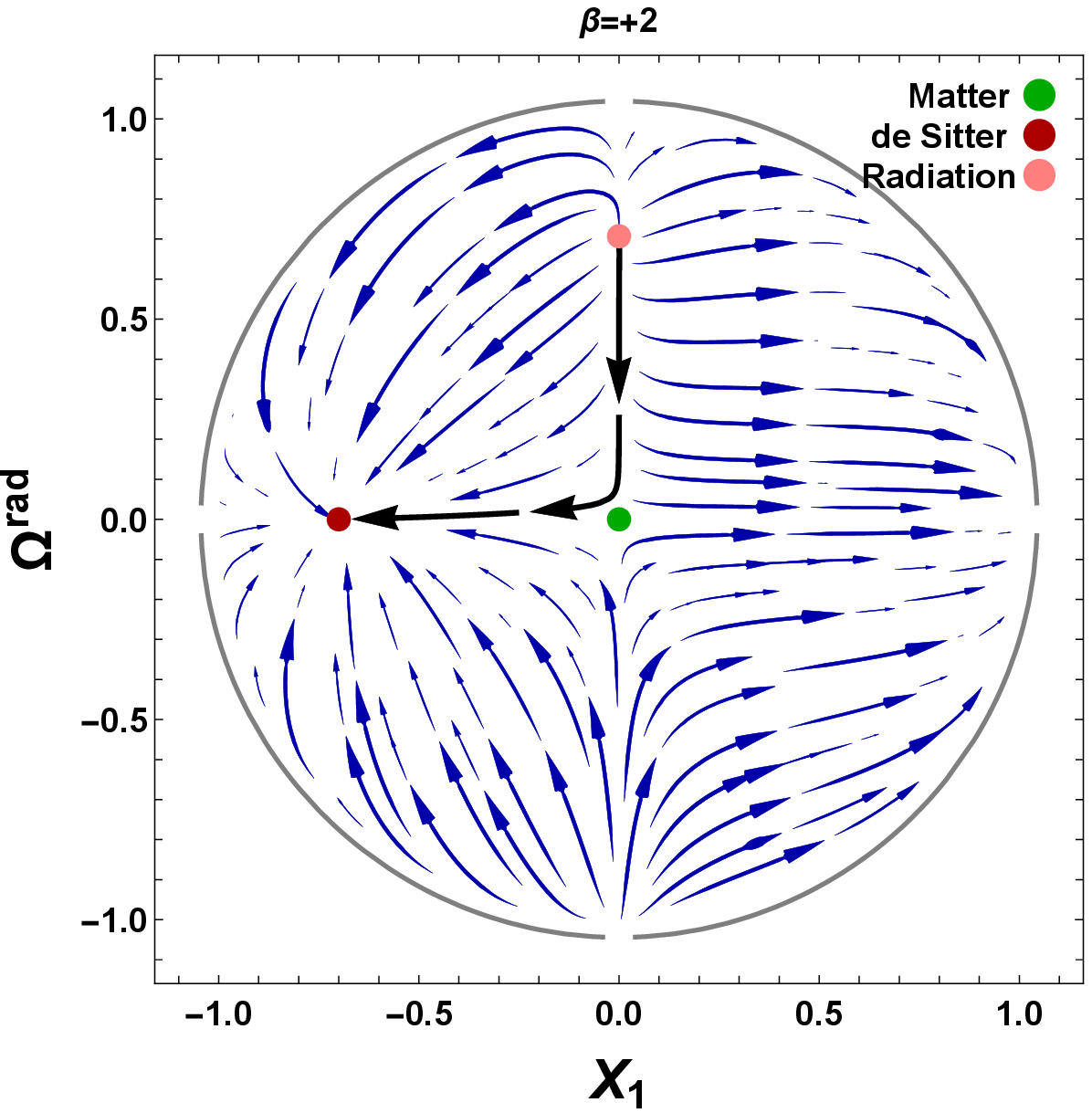,width=8.7cm}\hspace{2mm}
\epsfig{figure=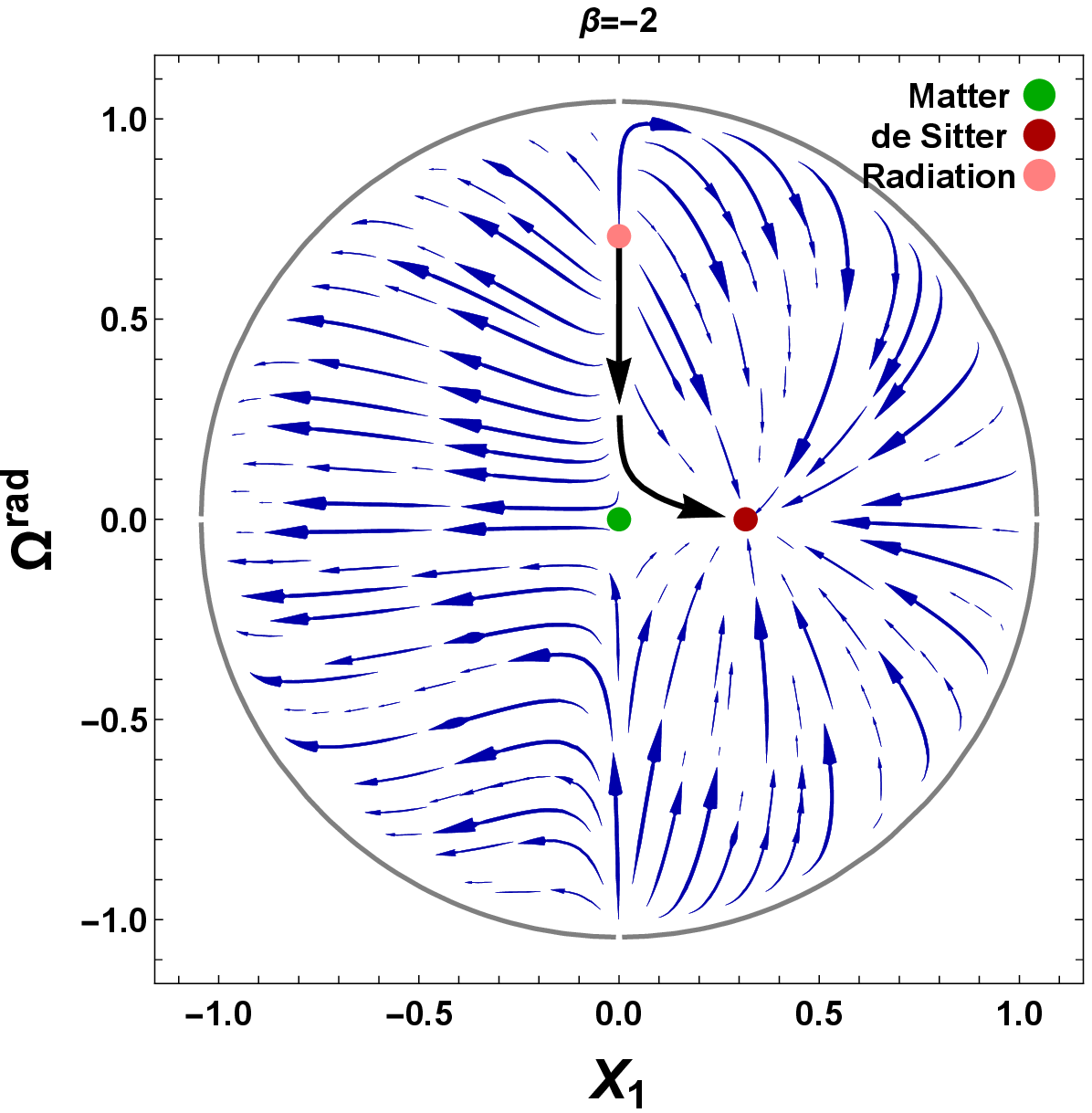,width=8.7cm}
\caption{Phase space portraits of $f(Q)=\alpha Q^{\beta}$ gravity for the case of constant $\gamma$ in the $x_3=0$ plane. }\label{fig3}
\end{center}
\end{figure}

In Fig.~\ref{fig3} we illustrate the power of $f(Q)$ gravity to make a true description of the ultra-relativistic to the dark matter transition which is followed by an everlasting accelerated expansion phase. Fig.~\ref{fig3} has been drawn in $x_3=0$ plane in which unphysical solutions $Q^a$-$Q^c$ do not appear. 

\begin{figure}[ht!]
\begin{center}
\epsfig{figure=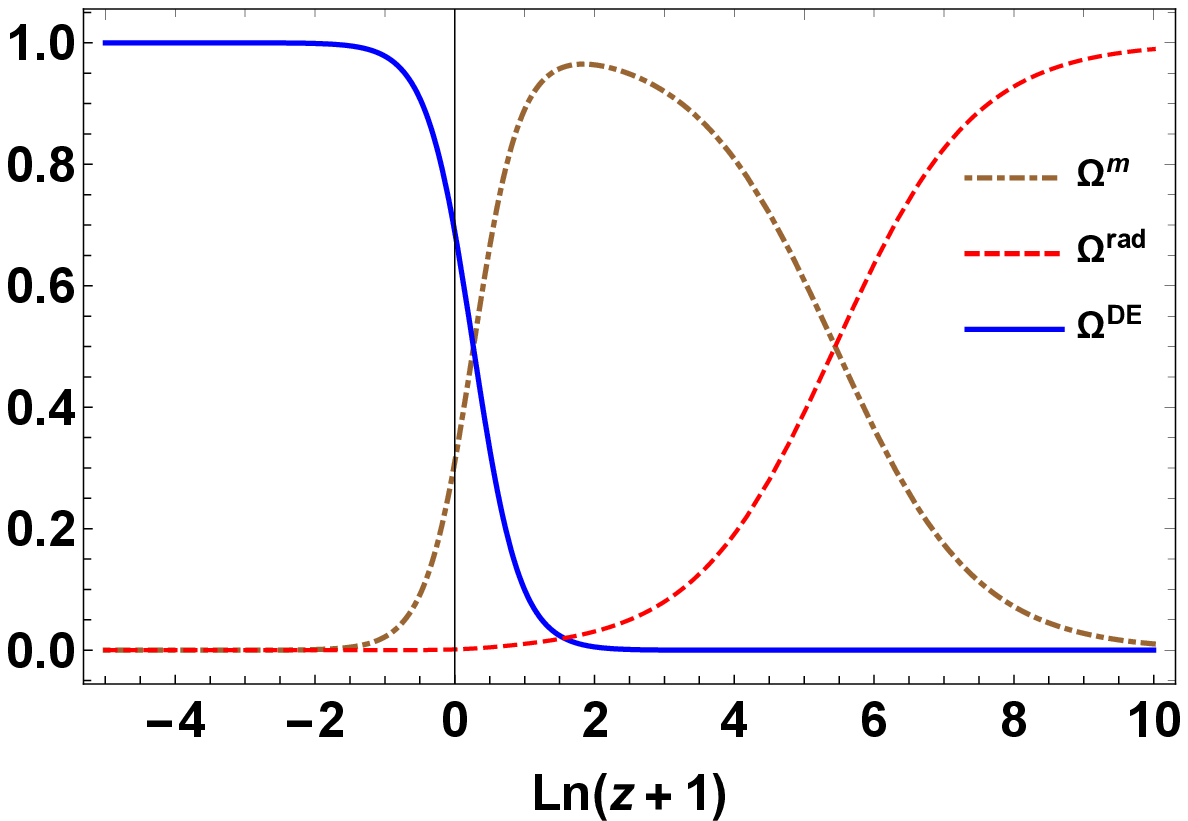,width=8.7cm}\hspace{2mm}
\epsfig{figure=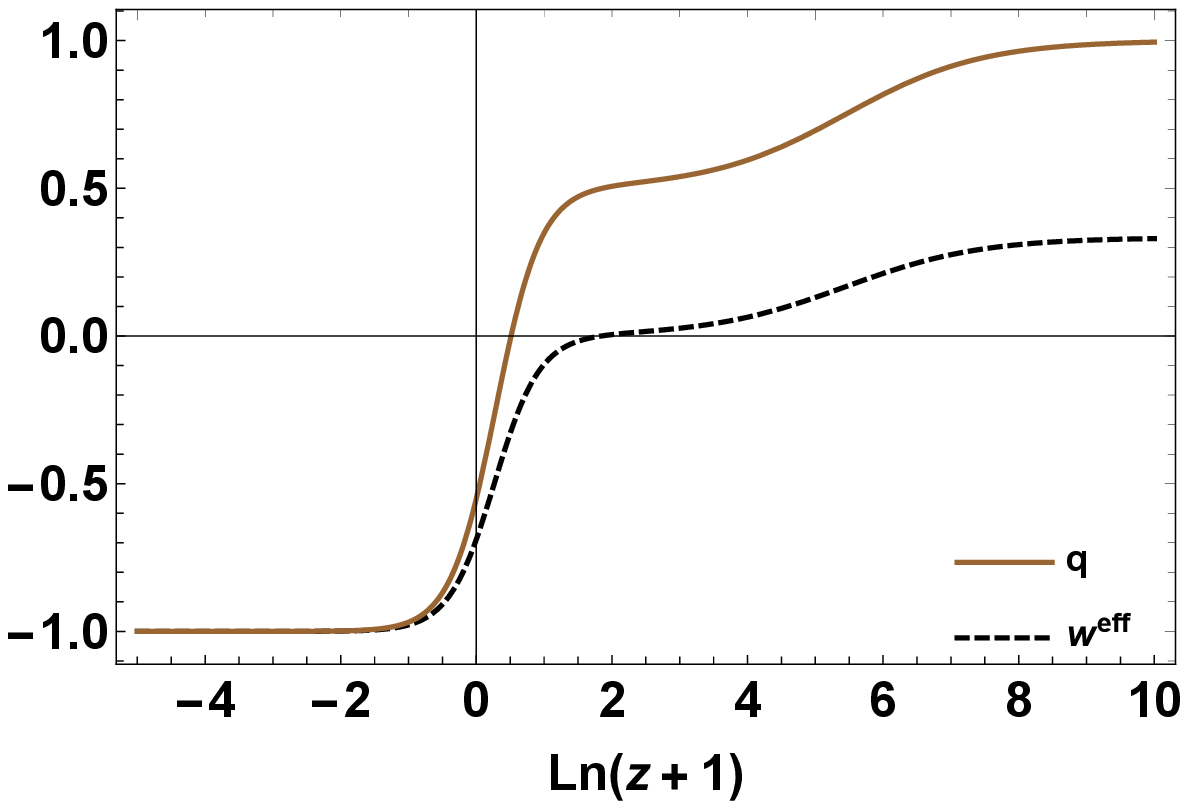,width=8.7cm}\vspace{2mm}
\epsfig{figure=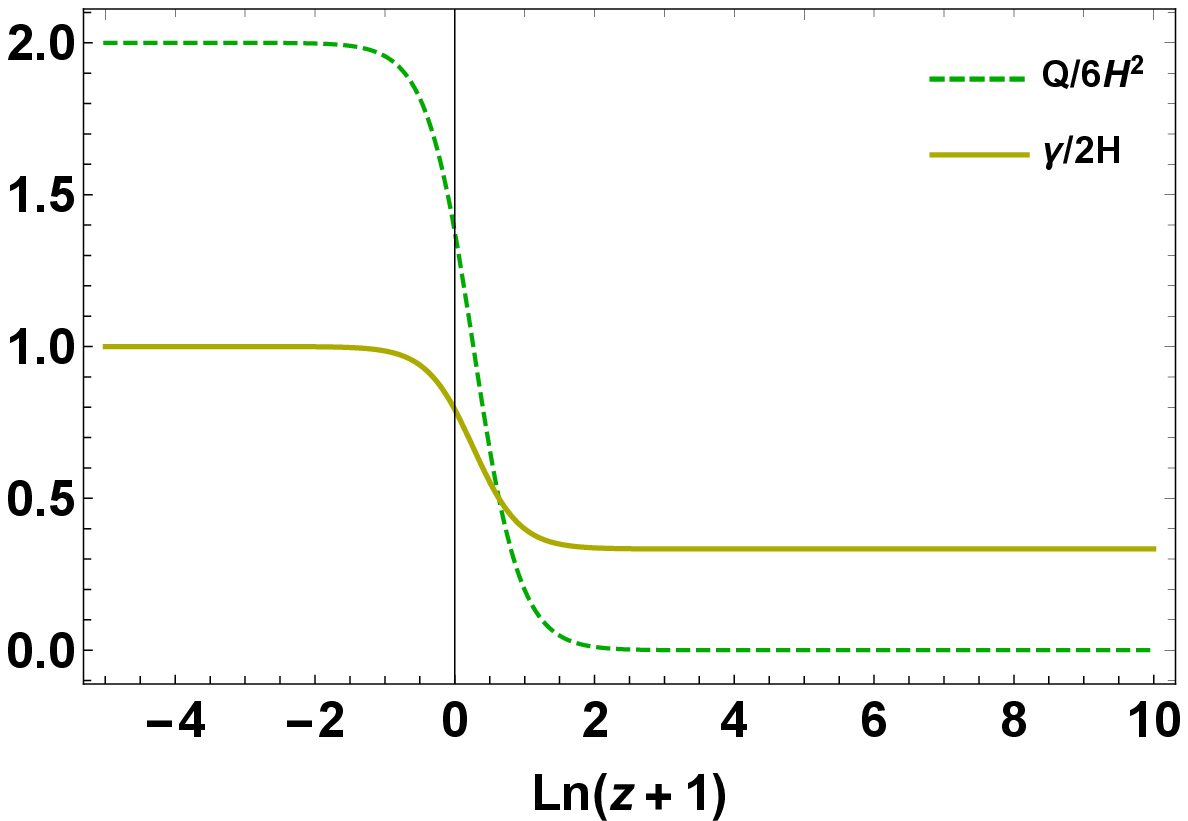,width=8.7cm}
\caption{Cosmological quantities of $f(Q)=\alpha Q^{\beta}$ gravity for $\beta=2$ in the case of constant $\gamma$. Initial values $x_{1i}=-1.8\times10^{-16}$, $x_{3i}=4.5\times10^{-14}$ and $\Omega^{rad}_i=0.999$ have been applied.}\label{fig4}
\end{center}
\end{figure}

Fig.~\ref{fig4} shows important cosmological quantities for $\beta=2$. These plots are made so that to give the present values $\Omega^m_0\approx0.31$ and $q_0\approx-0.55$ which are consistent with observations~\cite{planck2018}. We see the radiation-dark matter-accelerated expansion sequence from the above panels of Fig.~\ref{fig4}.

\subsection{$f(Q)=\eta  Q+\zeta  Q^{\sigma }$ theories of gravity}\label{cons-b} 
In these theories one has

\begin{align}\label{linear1}
m= \sigma\left(\frac{1}{r}+1\right) =\sigma\left(1 +\frac{ x_1}{x_2}\right).
\end{align}

Therefore, Eqs.~(\ref{con-g-41})-(\ref{con-g-43}) are closed by using expression~(\ref{linear1}). In this case, equation~(\ref{con-g-43}) stays unchanged while for Eqs.~(\ref{con-g-41}) and~(\ref{con-g-42}) we have

\begin{align}
&\frac{dx_1}{dN}= x_1 \Bigg[-3 (w+1) \left(x_1+x_2-1\right)-x_3 \Big((w-1) x_2+w\Big)\Bigg]+\frac{x_2^2 x_3}{\sigma(  x_1+ x_2)},\label{linear2}\\
&\frac{dx_2}{dN}=x_2 \Bigg\{x_2 \Bigg[x_3 \left(-w-\frac{1}{\sigma  (x_1+x_2)}+1\right)-3 (w+1)\Bigg]-3 (w+1) x_1-(w+1) \left(x_3-3\right)\Bigg\}.\label{linear3}
\end{align}
Solving the system of equations~(\ref{linear2}),~(\ref{linear3}) and~(\ref{con-g-43}), one finds that the de Sitter solution is the only physical one. In other words, the underlying model does not accept a solution corresponding to the dark matter dominated era. In this sense, these types of models cannot justify the evolution of the Universe in the first step.
\subsection{$f(Q)=\xi e^{\chi Q}$ theories of gravity}\label{cons-c} 

For the pure exponential models one gets $m=-r=-x_2/x_1$. Hence, in addition to Eq.~(\ref{con-g-43}) one obtains the following equations

\begin{align}
&\frac{dx_1}{dN}=x_1 \Bigg[-3 (w+1) \big(x_1+x_2-1\big)-x_3 \big((w-1) x_2+w+1\big)\Bigg],\label{exp1}\\
&\frac{dx_2}{dN}=x_1 x_3+x_2\Bigg[-3 (w+1) \left(x_1+x_2-1\right)-x_3 \Big((w-1) x_2+w+1\Big)\Bigg].\label{exp2}
\end{align}

 As can be seen the model parameters, i.e., $\xi$ and $\chi$ are not appeared in Eqs.~(\ref{exp1}) and~(\ref{exp2}) which signals that all possible results are valid independent of the values of these parameters. The effects of the model parameters can be extracted when one considers analytical solutions. Table~\ref{tabexp} explains the critical points of the system of Eqs.~(\ref{con-g-43}),~(\ref{exp1}) and~(\ref{exp2}) and their stabilities.

\begin{center}
\begin{table}[h!]
\centering
\caption{The fixed points solutions of $f(Q)=\xi e^{\chi Q}$ gravity with constant $\gamma$.}
\begin{tabular}{l @{\hskip 0.1in} l@{\hskip 0.1in} l @{\hskip 0.1in}l @{\hskip 0.1in}l}\hline\hline

Fixed point     &Coordinates $(x_1,x_2,x_3)$           &Eigenvalues  &$\Omega^m$ &$w^{eff}$\\[0.5 ex]
\hline
$R^{ds}_{e}$&$\left(0,0,3\right)$&$\left[0,0,-\frac{3}{2}  (w-1)\right]$&$0$&-1\\[1 ex]
$R^{m}$&$\left(0,0,0\right)$&$\left[\frac{3 (w-1)}{2},3 (w+1),3 (w+1)\right]$&$1$&0\\[1 ex]
$R^{ds}_{l}$&$\left(0,1,0\right)$&$\left[-3,0,-3 (w+1)\right]$&$0$&-1\\[0.75 ex]
$R^{ds}_{l*}$&$\left(1 - x_2,x_2,0\right)$&$\left[-3,0,-3 (w+1)\right]$&$0$&-1\\[0.75 ex]
\hline\hline
\end{tabular}
\label{tabexp}
\end{table}
\end{center}

\begin{figure}[ht!]
\begin{center}
\epsfig{figure=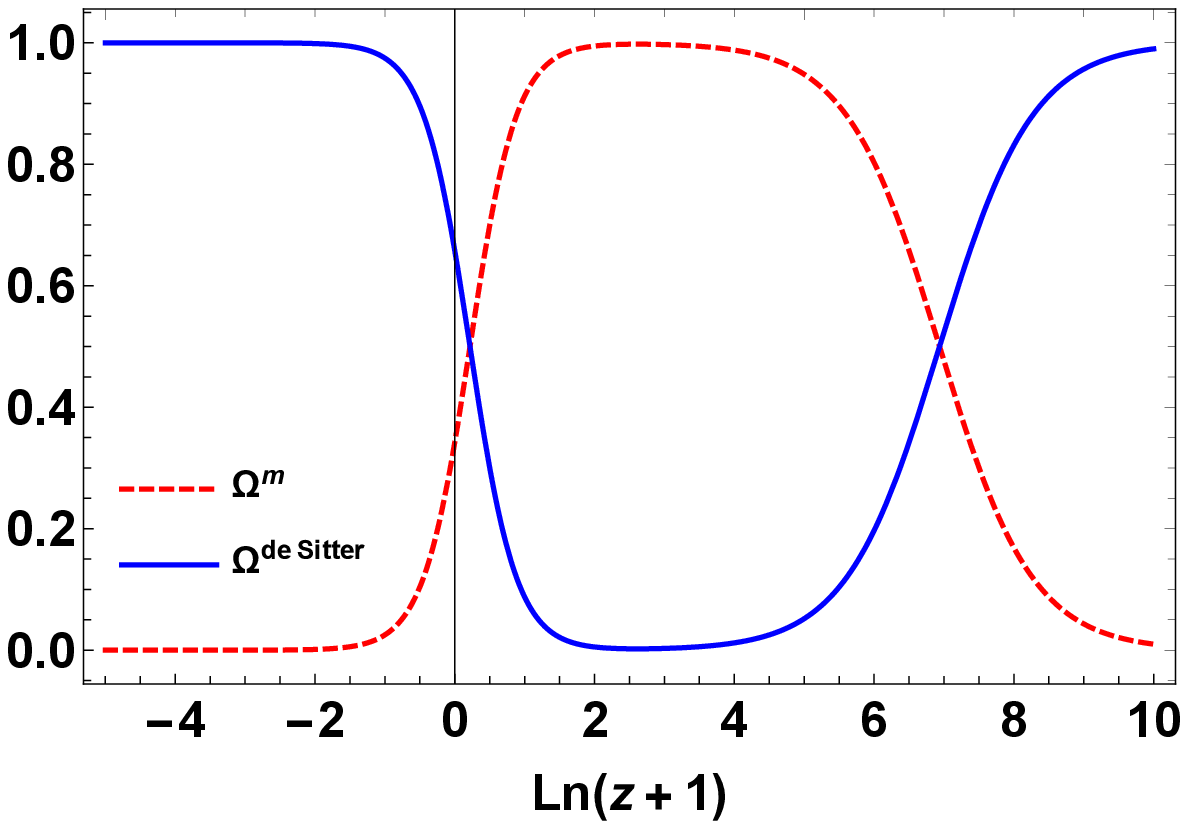,width=8.7cm}\hspace{2mm}
\epsfig{figure=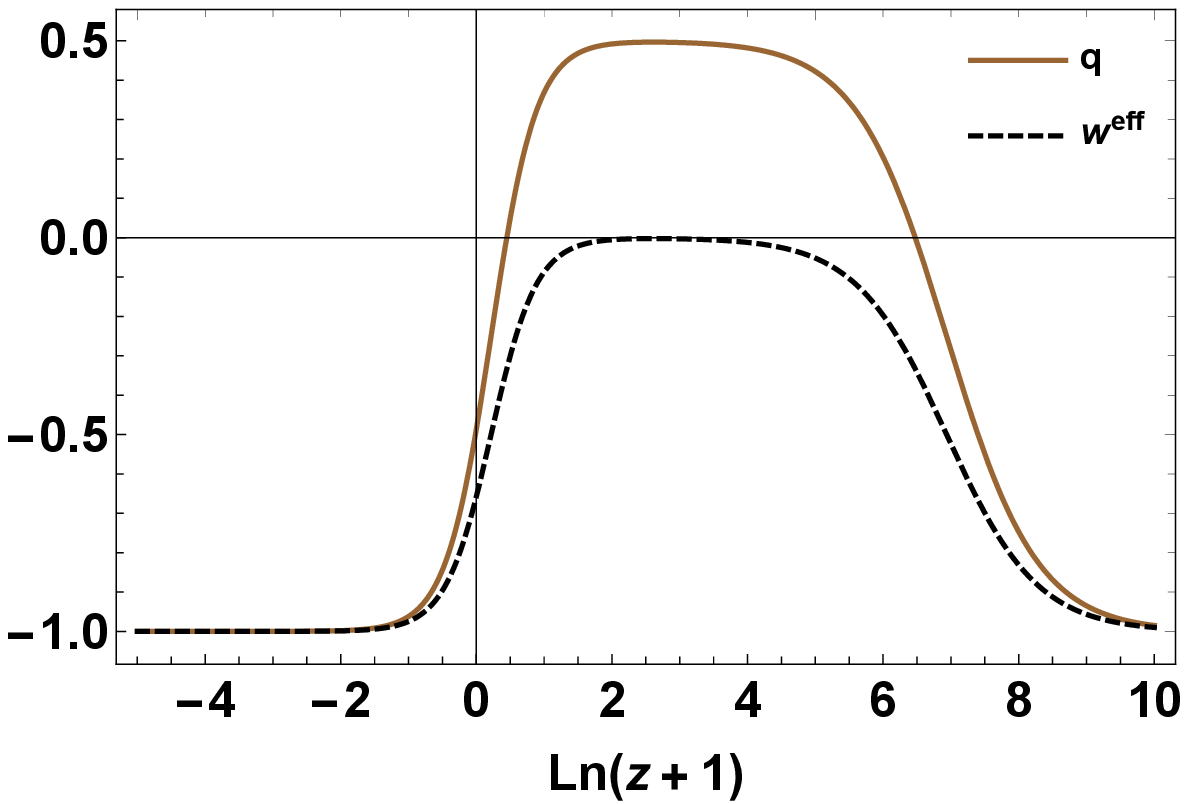,width=8.7cm}\vspace{2mm}
\epsfig{figure=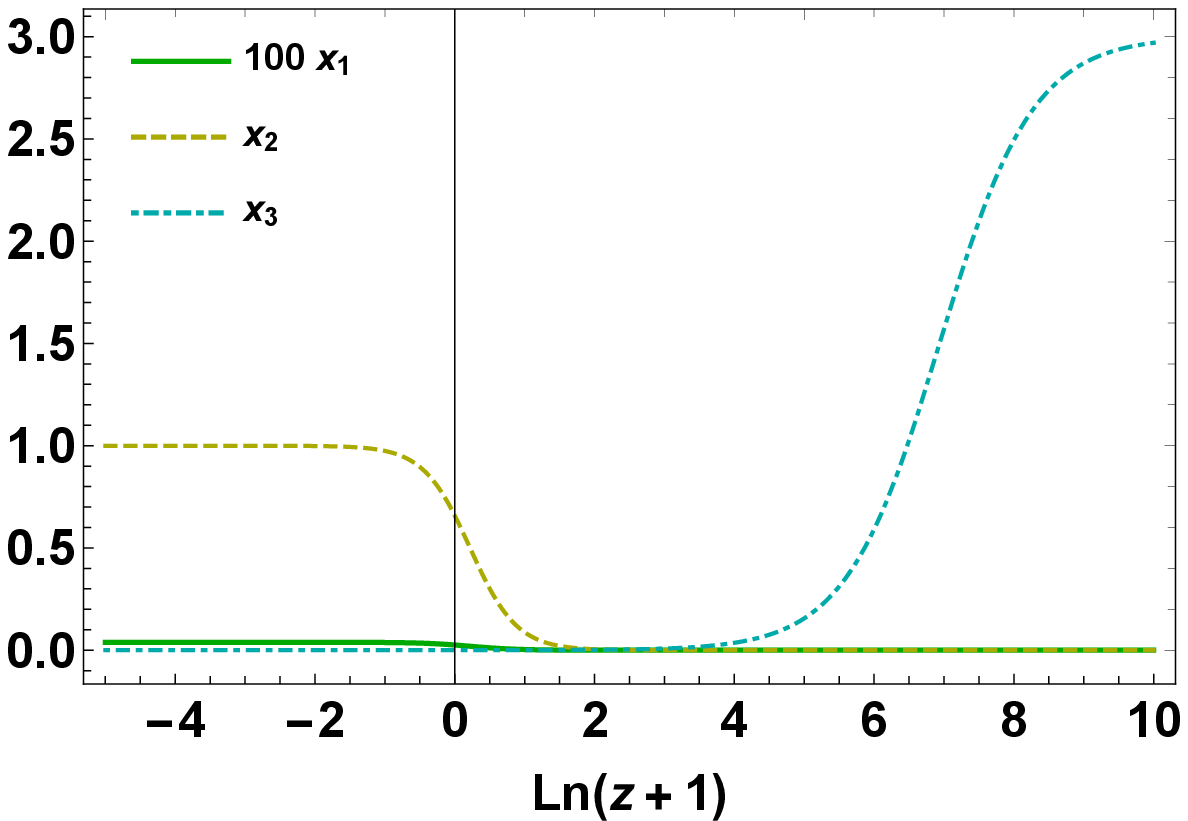,width=8.7cm}
\caption{Demonstration of the evolution of the Universe between early and late de Sitter eras in the $f(Q)=\xi e^{\chi Q}$ gravity theories. The initial values $x_{1i}=7 \times10^{-13}$, $x_{2i}=1.8\times10^{-9}$, $x_{3i}=2.97$ and $w=0$ have been set.}\label{figexp1}
\end{center}
\end{figure}

As can be seen, Table~\ref{tabexp} shows that besides the dark matter solution, $R^{m}$ and stable de Sitter solutions, $R^{ds}_{l}$ and $R^{ds}_{l*}$ (the former is a particular case of the latter), the pure exponential models accept an unstable de Sitter point for $w<1$. It seems very interesting since these type of models may also describe an early inflationary phase. Since, $R^{ds}_{l}$ and $R^{ds}_{l*}$ are attractive fixed points for $w>-1$, they only correspond to the late time accelerated expansion phase. Different panels in Fig.~\ref{figexp1} demonstrate the mentioned behaviors of de Sitter points. Particularly, the evolution of the phase space coordinates indicate a transition from $R^{ds}_{e}$ to $R^{ds}_{l}$. Note that, the smaller initial value, the less deviation of $x_1$ from being zero in the late times. 

\begin{figure}[ht!]
\begin{center}
\epsfig{figure=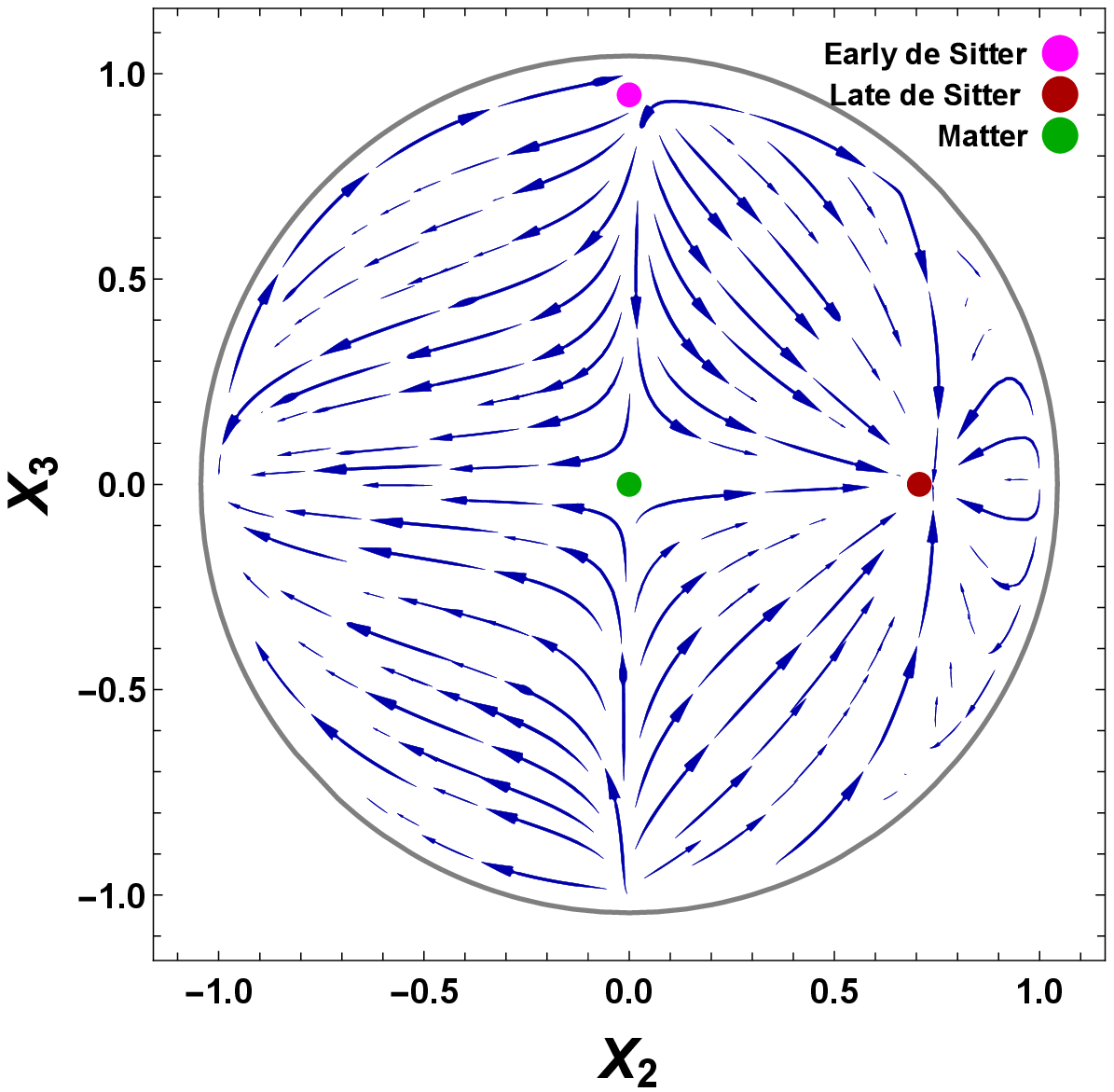,width=8.7cm}\hspace{2mm}
\epsfig{figure=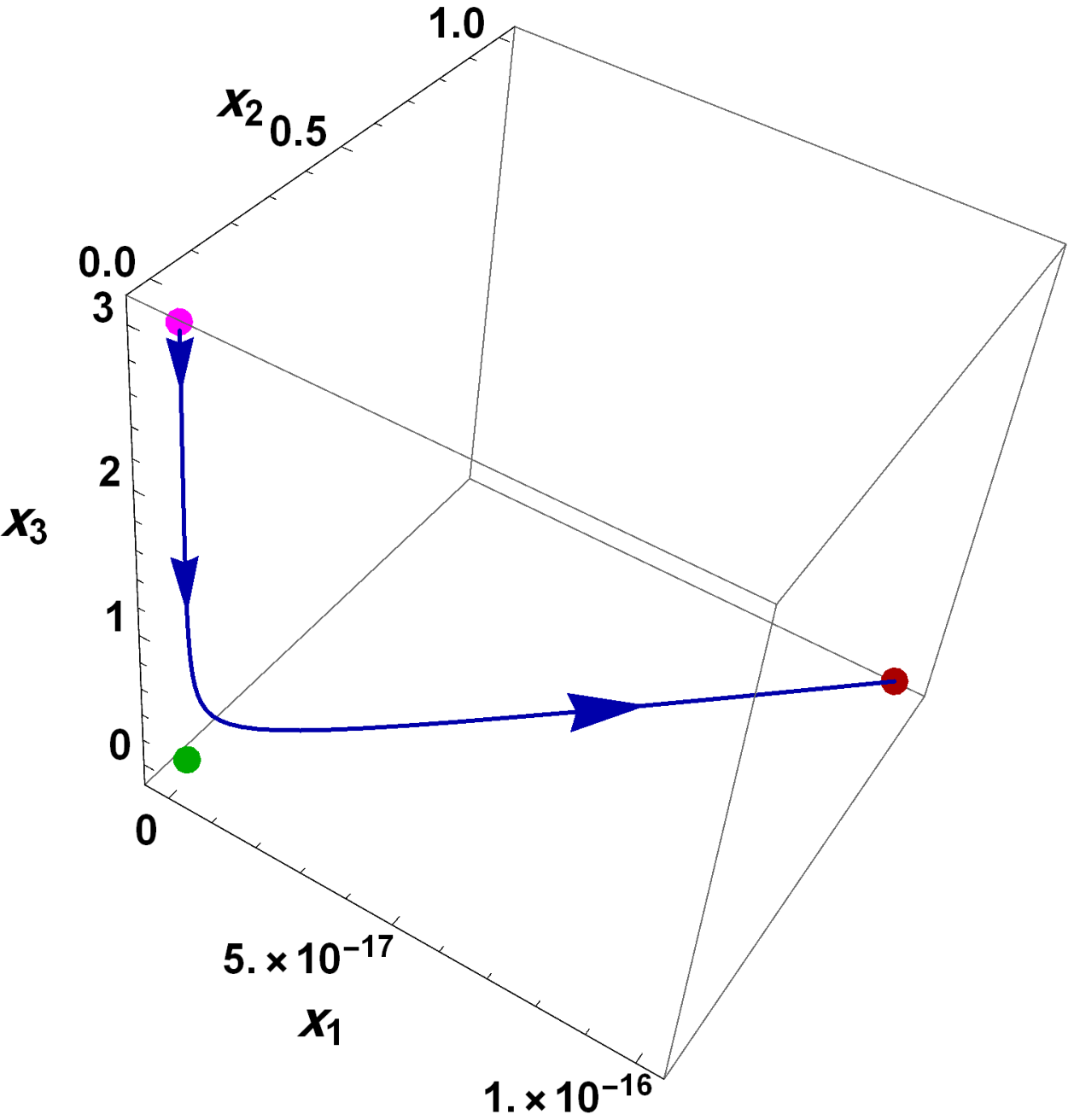,width=8.7cm}
\caption{Phase space illustration of Table~\ref{tabexp} for the initial values $x_{1i}=7 \times10^-20$, $x_{2i}=1.8\times10^{-4}$, $x_{3i}=2.97$.}\label{figexp2}
\end{center}
\end{figure}

Also, in Fig.~\ref{figexp2} we have drawn the phase portraits in two and three dimensional phase spaces. The existence of two de Sitter fixed points with right stability conditions is very exciting since this result has been obtained without employing any extra scalar field which usually is done in inflationary theories.

\section{Concluding remarks}\label{sec8}

In symmetric teleparallel equivalent of general relativity, non-metricity characterises the gravity in a curvature free and torsion free environment. We have studied a natural extension of it, the modified $f(Q)$ theory which has been proved to be successful in demonstrating the late time acceleration of the Universe without assumption of dark energy. In the present study, we have focused particularly into the cosmological side of this theory. It has been noticed that so far the researchers only utilised the coincident gauge formulation (considering a vanishing affine connection) while discussing the cosmological application of $f(Q)$ theory in a spatially flat FLRW spacetime. This gauge choice has a serious limitation, as the obtained Friedmann type equations match exactly with those of the well-studied torsion-based $f(\mathbb{T})$ theory. So unknowingly, researchers have been reproducing the known results masked in a new packaging, it seems. 

In our current study therefore we have formulated a fresh new $f(Q)$ theory dynamics using a class of non-vanishing affine connection involving a free parameter $\gamma(t)$, whose Friedmann type equations are completely aloof from $f(\mathbb{T})$ theory. We have derived the field equations, the pressure and energy density equations for this novel construction. The covariant divergence of the stress-energy tensor has been derived, and in generic $f(Q)$ models it has been displaying non-zero components.

To analyse the complicated system, we have used the dynamical system analysis method. To construct an inter-related dimensionless system, we have first assumed the compatibility of the metric field equation with energy conservation which has provided us with an additional constraint. Two different assumptions have been studied. We firstly have considered the theory with a general time-varying $\gamma(t)$ and then with a time-independent one. In the former case the model $f(Q)=\alpha Q^{\beta}$ has been investigated. A stable critical point which describes a de Sitter era and an unstable one representing a matter dominated phase have been found. In addition to the mentioned solutions there also exist some stable fixed points implying accelerated expansion phases with $w^{eff}<-1/3$ which cannot be connected to the matter-dominated point in the phase space. In this regard, we have called such solutions as isolated fixed points. Hence, a de Sitter point is the only attractor of the phase space trajectories emanating from the matter fixed point.

Next, the $f(Q)$ gravity theory with time-independent $\gamma$ has been studied. Similar results have been obtained. As a particular case we have considered the $f(Q)=\alpha Q^{\beta}$ model when both the ultra-relativistic matter and the dark matter are present. An acceptable track of radiation-dark matter-dark energy dominated epochs can be pictured in this model.  The models with $f(Q)=\eta  Q+\zeta  Q^{\sigma }$ suffer from lacking a matter dominated solution. We have also analyzed the models with $f(Q)=\xi e^{\chi Q}$ function and found that these theories describe the early times de Sitter era in addition to the late times one. Accordingly, the pure exponential models justify the transitions de Sitter-matter dominated-de Sitter eras.

In summary, $f(Q)$ gravity theories are capable of describing de Sitter expansions both in the early and the late times with a middle matter dominated phase, depending on the form of $f(Q)$ function. Within the frame of the dynamical system, one only can inspect the model for existing desirable solutions. Further details including constraining model parameters using the astronomical data should be followed by analytically solving the field equations which is our next program of study.


\end{document}